\documentstyle[aps]{revtex}
\begin{document}
\title{Phenomenological predictions of the properties of the $B_{c}$ system}
\draft
\author{Lewis P. Fulcher}
\address{Department of Physics and Astronomy, Bowling Green State University\\
Bowling Green, Ohio 43403\\ email: fulcher@newton.bgsu.edu\\ and\\
Institut f\"{u}r Theoretische Physik der Justus Liebig Universit\"{a}t\\
D-35392 Gie\ss en, Germany }
\maketitle
\begin{abstract}
We present a comprehensive calculation of the energies, splittings and
electromagnetic decays of the low-lying levels of the bottom charmed 
meson system. In order to incorporate running coupling constant effects,
we choose Richardson's potential for the central potential and take the
spin-dependent potentials from the radiative one-loop calculation of
Pantaleone, Tye and Ng. The effects of a nonperturbative spin-orbit
potential are also included. Our parameters are determined from the
low-lying levels of charmonium (avg. dev. of 19.9 MeV) and of the
upsilon system (avg. dev. of 4.3 MeV). We carry out a detailed comparison
with the calculations of Eichten and Quigg and the lattice calculations
of the NRQCD collaboration. Our predicted result for the ground state energy is
$6286_{-6}^{+15}$ MeV. Our results are generally in agreement with the
earlier calculations. However, we find the two lowest $1^{+}$ states to be very
close to the j-j limit, in contrast to some of the earlier calculations.
The implications of this finding for the photon spectra of the 1P and
2S states are discussed in detail. Some strategies for the observation
of these states are discussed, and a table of their cascades to the
ground state is presented. Our calculated value for the ground state
lifetime is $0.38 \pm 0.03 $ ps, in good agreement with the recent CDF
measurement.
\end{abstract}

\section{introduction}
\label{sec:in}
The CDF collaboration \cite{ab98} has reported the discovery
of the $B_{c}$ system in 1.8 TeV $p-\bar{p} $ collisions at Fermilab. They
have observed about 20 decays in the $J/\psi $-lepton channel, 
which are interpreted as decays of the ground state. For the mass of the
ground state, the CDF collaboration quotes $ M_{B_{c}} = 6.40 \pm 0.39 \pm
 0.13 \; {\rm GeV} $. Their value for the lifetime is $\tau_{B_{c}} =
0.46^{+0.18}_{-0.16} \pm 0.03$ ps. This state should be one of
a number of states lying below the threshold for emission of B and D mesons.
Because these states can not decay by gluon annihilation, they should be very
stable in comparison with their counterparts in charmonium and the upsilon
system. 

The purpose of this report is to give a detailed account of the
energies, splittings and electromagnetic decay rates for the $B_{c}$ states 
below the continuum threshold and to propose strategies for detecting
some of these states. We will use a potential model that includes running
coupling constant effects in both the central potential and the spin-dependent
potentials to give a simultaneous account of the properties of
charmonium, the upsilon system and the $B_{c}$ system. We choose
Richardson's potential \cite{ri79} to represent the central potential and
insist upon strict flavor-independence of its parameters. Since one
would expect the average values of the momentum transfer in the various
quark-antiquark states to be different, some variation
in the values of the strong coupling constant and the renormalization
scale in the spin-dependent potentials should be expected. 
In order to minimize the role of flavor-dependence, we use the same
values for the coupling constant and the renormalization scale for each of the
levels in a given system and require that these values 
be consistent with a universal QCD scale. Since one can calculate
the central potential and the spin-dependent potentials from first
principles on the lattice \cite{ba97}, it should be possible
at some point in future calculations to connect subsets of these potential
parameters in more fundamental ways. Such a program would be an important
step towards finding a more rigorous way to formulate a potential model
calculation.

In 1991 Kwong and Rosner \cite{kw91} predicted the masses of the lowest
vector and pseudoscalar states of the $B_{s}$ and the $B_{c}$ systems 
using an empirical mass formula and a logarithmic potential. Eichten
and Quigg \cite{ei94} gave a more comprehensive account of the energies 
and decays of the of the $B_{c}$ system that was based on the 
QCD-motivated potential of Buchm\"{u}ller and Tye \cite{bu81}. Gershtein
{\em et al.} \cite{ge95} also published a detailed account of the energies
and decays of the $B_{c}$ system and established contact with QCD 
sum-rule calculations. Both of these latter calculations included running
coupling constant effects in the central potential, but used spin-dependent
potentials that were restricted to the tree level. 

One of the most important goals of the present calculation is to extend
the treatment of the spin-dependent potentials to the full radiative
one-loop level and thus include effects of the running coupling constant
in these potentials. In a previous calculation \cite{fu91} we have shown
that such effects offer a substantial improvement in the calculation of
the spectra of charmonium and the upsilon system. In particular, it
becomes possible to offer a good account of both the fine structure
splittings and the hyperfine structure. Our previous calculation
used the renormalization scheme developed by Gupta
and Radford \cite{gu81,gu82}. However for this calculation, we have chosen the
modified-minimal subtraction scheme used by Pantaleone, Tye and Ng (PTN)
\cite{pa86} to extend their perturbative QCD calculation to the one-loop level.
We supplement the PTN formalism with a long-range spin-orbit potential
to be consistent with the Gromes consistency condition \cite{gr84} and recent
lattice calculations \cite{ba97}.
The flavor dependence allowed the string constant in the
1991 calculation is not permitted in the present calculation. Thus this 
calculation should also be viewed as part of the effort to formulate
potential model calculations under a more restrictive set of assumptions.
We note in passing
a number of additional important calculations of the properties of the
$B_{c}$ system \cite{ei81,go85,ba92,ch92,ga92}.

In Sect. \ref{sec:so} we present preliminary calculations of 
the lowest vector and pseudoscalar states of the $B_{c}$ system with
several potential models. These exercises are similar to some carried out by
Eichten and Quigg \cite{ei94} and represent one way of estimating the 
uncertaintities in the predicted energies of the new system. In Sect.
\ref{sec:ri} we introduce Richardson's potential and the expressions necessary 
to include the one-loop corrections in the spin-dependent potentials and
discuss the determination of our parameters from charmonium and the upsilon
system. The formalism required for the mixing of the P states in the $B_{c}$
system is presented in Sect. \ref{sec:ma}. The last section contains our
results for the energies and decays of the low-lying $B_{c} $ states and 
a discussion of their implications for the development of strategies to
detect these states.

\section{some potential models}
\label{sec:so}

The three systems that we consider in this paper, upsilon, charmonium and       $B_{c}$, are often considered as nonrelativistic systems, and thus our
treatment is
based upon the Schr\"{o}dinger equation with a Hamiltonian of the form,

\begin{equation}
\label{nrham}
H = p^{2}/2\mu + V(r) + V_{\rm SD},
\end{equation}
where $ \mu = m_{1} m_{2} /(m_{1} + m_{2})$ is the reduced mass. No 
spin-independent relativistic corrections are included. For the purpose
of making some
preliminary estimates of the energies of the lowest two states of the $B_{c}$
system, it is necessary to consider only the spin-spin part of the 
spin-dependent potential since these are S states. Thus,

\begin{equation}
\label{conpot}
V_{SD} \rightarrow V_{SS} = \frac{32 \pi \alpha_{S}}{9 m_{1} m_{2}} 
	\delta^{3} ( {\bf r}) {\bf S_{\rm 1} \cdot S_{\rm 2}}.
\end{equation}
Our solutions to the Schr\"{o}dinger equation are generated numerically with
the given form for the central potential \cite{fu88} and the effects of the
spin-dependent parts are added as a perturbation improvement. Thus, the 
1S-state hyperfine splitting is given by

\begin{equation}
\label{hfs}
\Delta E_{\rm hfs} = \frac{8 \alpha_{S} }{9 m_{1} m_{2}} |R_{1S}(0)|^{2}.
\end{equation}
We address the question of the validity of first-order perturbation theory
for the contact potential of Eq. (\ref{conpot}) in an Appendix.

All of the potential parameters in this section are strictly 
flavor-independent. This includes any additive constants. The central potential
parameters and the constituent mass parameters are fit to the low-lying energy
levels of charmonium and the upsilon system. The strong coupling constant
$\alpha_{S}$ is fit to the observed charmonium hyperfine splitting of 117 MeV.
In Table I below we present results with 5 different central potentials. 
These include the Cornell potential \cite{ei78}

\begin{equation}
V(r) = Ar - \kappa /r + C,
\end{equation}
where $A=0.1756 \; {\rm GeV^{2}} $, $\kappa = 0.52$ and $C = -0.8578 GeV $;
Mart\'{\i }n's power-law potential \cite{ma80},

\begin{equation}
V(r) = - 8.093 + 6.898 \: (r \times {\rm 1 GeV} )^{0.1} ,
\end{equation}
(units of the potential are GeV), and the logarithmic potential \cite{qu77},

\begin{equation}
\label{vlog}
V(r) = -0.6631 + 0.733 \ln (r \times {\rm 1 GeV}) ,
\end{equation}
(potential units are also GeV). 
Each of these forms was used by Eichten and Quigg. However, our values for the
potential parameters are slightly different. We also have done calculations
with the potential of Song and Lin \cite{so87},

\begin{equation}
V(r) = A r^{1/2} -B r^{-1/2},
\end{equation}
where $A = 0.511 \; {\rm GeV^{3/2}}$ and $B = 0.923 \; {\rm GeV^{1/2}}$, and
the Turin \cite{li89} potential,

\begin{equation}
V(r) = - a r^{-3/4} + b r^{3/4} + C,
\end{equation}
where $a = 0.620 \; {\rm GeV^{1/4} }$, $ b = 0.304 \; {\rm GeV^{7/4}}$ and
$ C = -0.823 \; {\rm GeV}$.

Our results in the first 3 columns of Table \ref{tab1} are very similar to 
those presented in Table I of Eichten and Quigg \cite{ei94}. The results
obtained with the Song-Lin and the Turin potentials in all cases fall
between the extremes defined by the first 3 columns. Averaging over all 5
cases presented in Table \ref{tab1} yields

\begin{equation}
M_{B_{c}} = 6251^{+15}_{-5} \; {\rm MeV}, \hspace{0.5in} M_{B^{*}_{c}} =
      6328^{+9}_{-8} \; {\rm MeV},
\end{equation}
as the predicted energies of the two lowest states.
If we treat Eichten and Quigg's results in the same way, then we get
$ 6258^{+8}_{-10} \; {\rm MeV}$ for $B_{c}$ and $ 6333^{+10}_{-14} \; 
{\rm MeV} $
 for $B_{c}^{*}$,
in good agreement with our preliminary analysis.
All of these results fall comfortably within the ranges found by Kwong
and Rosner \cite{kw91}, that is,

\begin{mathletters}
\begin{equation}
6194 \; {\rm MeV} \; \leq \; M_{B_{c}} \; \leq 6292 \; {\rm MeV}, 
\end{equation}
\begin{equation}
6284 \; {\rm MeV} \; \leq \; M_{B^{*}_{c}} \; \leq 6357 \; {\rm MeV} .
\end{equation}
\end{mathletters}

\section{Richardson's potential and the spin-dependent potentials}
\label{sec:ri}

The starting point for the derivation \cite{lu91}
of Richardson's potential is the 
one-loop expression for the running coupling constant \cite{gr87} , namely,

\begin{equation}
\label{ru}
\alpha_{S}(|{\bf q}^{2}|) = \frac{ 12 \pi }{ (33 - 2 n_{f}) \ln (
           |{\bf q}^{2}|/\Lambda^{2} ) },
\end{equation}
where $n_{f}$ denotes the number of quark degrees of freedom accessible to
the propagating gluon \cite{ft1} and $\Lambda $ is the QCD scale. 
Richardson \cite{ri79} realized that one could tame the infrared
singularity of Eq. (\ref{ru}) and produce a linear confining
potential by making the replacement,

\begin{equation}
|{\bf q}^{2}|/\Lambda^{2} \: \rightarrow \: |{\bf q}^{2}|/\Lambda^{2} + 1 ,
\end{equation}
in the argument of the logarithm. Our 1991 calculation \cite{fu91} 
was based on a modification of Richardson's potential
suggested by Moxhay and Rosner \cite{mo83}, that is, the string constant
is treated as a free parameter instead of being connected to the
scale parameter $\Lambda $. Then the potential takes the form,

\begin{equation}
V(r) = A r - \: \frac{8\pi \, f(\Lambda r ) }{(33 - 2 n_{f})r} ,
\end{equation}
where
\begin{equation}
f(t) = \frac{4}{\pi } \int_{0}^{\infty } \frac{sin\: tx}{x} \left[
\frac{1}{\ln (1 + x^{2})} - \frac{1}{x^{2}} \right] dx.
\end{equation}

To specify the spin-dependent potentials, we use a generalization
of the Eichten-Feinberg fromalism \cite{ei81} derived by PTN \cite{pa86}.
Our notation is closely akin to that of Eichten
and Quigg \cite{ei94}. Thus,

\begin{equation}
\label{sd}
V_{SD} = \frac{{\bf L} \cdot {\bf S}_{1} } { 2 m_{1}^{2} } T_{a}
       + \frac{{\bf L} \cdot {\bf S}_{2} } { 2 m_{2}^{2} } T_{a}^{\prime }
  + \frac{{\bf L} \cdot ({\bf S}_{1} + {\bf S}_{2}) } { m_{1} m_{2} } T_{b}
+ \frac{{\bf L} \cdot ({\bf S}_{1} - {\bf S}_{2}) }{m_{1} m_{2}}T_{b}^{\prime}
+ \frac{S_{12}}{m_{1} m_{2}} T_{c} 
+ \frac{ {\bf S}_{1} \cdot {\bf S}_{2} } {m_{1} m_{2} } T_{d}, 
\end{equation}
where $S_{12}$ denotes the tensor operator, that is, $ 4 \left[ 3 {\bf S}_{1}
\cdot \hat{\bf r} {\bf S}_{2} \cdot \hat{\bf r} - {\bf S}_{1} \cdot {\bf S}_{2}
\right] $ and $m_{1}$ denotes the mass of the lighter (charmed) quark, if the
two constituent masses are not equal. The quantities T are connected with
the PTN potentials as follows:

\begin{mathletters}
\begin{equation}
T_{a} = \frac{1}{r} \frac{d}{d\,r} ( E - Ar) + \frac{2}{r} \frac{d\, V_{1}}
        {d\,r} + 2 V_{5}, \hspace{0.2in} T_{a}^{\prime } = T_{a} - 4 V_{5},
\end{equation}
\begin{equation}
T_{b} = \frac{1}{r} \frac{d\, V_{2} }{d \, r}, \hspace{0.2in} T_{b}^{\prime }
         =V_{5},\hspace{0.2in} T_{c} = V_{3}/12,\hspace{0.2in} T_{d} = V_{4}/3,
\end{equation}
\end{mathletters}
where we have supplemented the PTN potentials with a long range spin-orbit
contribution to include nonperturbative effects. Such a contribution is 
clearly indicated by the lattice calculations of Bali, Schilling and 
Wachter \cite{ba97} and is required to satisfy the consistency condition,
\begin{equation}
\frac{d }{d \,r} \left[ E(r) + V_{1}(r) -V_{2}(r) \right] = 0,
\end{equation}
derived by Gromes \cite{gr84} from the requirements of Lorentz covariance.

The PTN potentials are expressed in terms of the strong coupling constant
$\alpha_{\bar{S}}$ and the renormalization scale $\mu $. The function E(r)
includes the leading correction to the central potential,

\begin{equation}
E(r) = - \frac{4 \alpha_{\bar{S}}}{3 r} \left[ 1 + \frac{\alpha_{\bar{S}}}
      {6 \pi } \left[ (33 -2 n_{f}) (\ln \mu r + \gamma_{E} ) + \frac{31}
      {2} - \frac{5 n_{f} }{3} \right] \right],
\end{equation}
where $\gamma_{E} = 0.5772 \cdots $, is Euler's constant. This function is
of course a part of the contribution to the short-distance behavior of the
running coupling constant in Eq. (\ref{ru}), although its contribution to
Richardson's potential is not made explicit.
The first two spin-orbit potentials are given by

\begin{equation}
V_{1}(r) = - \frac{ \alpha_{\bar{S}}^{2}} {2 \pi r} \left[ \frac{16}{9} -4 (
     \ln \sqrt{m_{1} m_{2}} r  + \gamma_{E} ) \right],
\end{equation}

\begin{equation}
V_{2}(r) = - \frac{ 4 \alpha_{\bar{S}}}{3 r} \left[ 1 + \frac{\alpha_{\bar{S}}}
      { 6\pi }\left[ (33 - 2 n_{f} ) (\ln \mu r + \gamma_{e} ) + \frac{39}{2}
      - \frac{5 n_{f} }{3} - 9 ( \ln \sqrt{m_{1} m_{2}} r + \gamma_{E} )
      \right] \right].
\end{equation}
The tensor potential takes the form,

\begin{equation}
V_{3}(r) =\frac{ 4 \alpha_{\bar{S}}}{r^{3}} \left[ 1 + \frac{\alpha_{\bar{S}}}
      {6 \pi } \left[ (33 - 2 n_{f}) ( \ln \mu r + \gamma_{E} - \frac{4}{3})
      +\frac{65}{2} - \frac{ 5 n_{f}}{3} -18 ( \ln \sqrt{m_{1} m_{2}} r +
      \gamma_{E} - \frac{4}{3}) \right] \right].
\end{equation}
The lengthy expression for the spin-spin potential,

\begin{eqnarray}
\label{ss}
V_{4}(r) & = & \frac{ 32 \pi \alpha_{\bar{S}} }{3} \left\{  1 +
      \frac{\alpha_{S} }{ 6 \pi} \left[ \frac{11}{2} - \frac{ 5 n_{f}}{3}
     - \frac{3}{4} \left( \frac{ 9 m_{1}^{2} + 9 m_{2}^{2} -14 m_{1} m_{2}}
       {m_{1}^{2} -m_{2}^{2}} \right) \ln (m_{2}/m_{1})
  \right] \right\} \delta^{3} ({\bf r})  \nonumber \\
    & & - \frac{4 \alpha_{\bar{S}}^{2}}{9 \pi }
 \left[ (33 -2n_{f}) \nabla^{2} \left( \frac{\ln \mu r + \gamma_{E}}{r} \right)
  - \frac{63}{2} \nabla^{2} \left( \frac{ \ln \sqrt{m_{1} m_{2} } r + 
      \gamma_{E} }{r} \right)
        \right]  ,
\end{eqnarray}
reduces to an especially simple form for states with nonzero angular
momentum, that is,
\begin{equation}
V_{4} \rightarrow \frac{ \alpha_{\bar{S}}^{2} }{ \pi r^{3}} \left[
    \frac{2}{3} - \frac{8 n_{f}}{9} \right],
\end{equation}
since one can take the derivatives present in Eq. (\ref{ss}) 
in this case, and the delta function there does not contribute.
If the two constituent quarks
have equal masses, then it is necessary to add a second-order
contribution from the annihilation graphs,
\begin{equation}
\delta V_{4} = 8 \alpha_{\bar{S}}^{2} ( 1 - \ln 2 ) \delta^{3} ({\bf r} ),
\end{equation}
to the expression of Eq. (\ref{ss}). The spin-orbit potential discovered
by PTN is given by
\begin{equation}
V_{5} (r) = \frac{\alpha_{\bar{S}}^{2} } {\pi r^{3} } \ln (m_{1}/m_{2}).
\end{equation}

The equations describing the PTN potentials use the modified minimal subtraction
scheme to define the strong coupling constant $\alpha_{\bar{S}} $. This coupling
constant can be related to that defined in a scheme developed by Gupta 
{\em et al. } \cite{gu81,gu82}, which was used in the 1991 calculation 
\cite{fu91}, that is,

\begin{equation}
\label{tr}
\alpha_{GRR}(\mu^{2}) = \alpha_{\bar{S}}(\mu^{2}) \left[ 1 + \frac{
    \alpha_{\bar{S}}(\mu^{2}) }{ 12 \pi } \left( 49 - \frac{10 n_{f}}{3}
    \right) \right] + \cdots ,
\end{equation}
to the second order in $\alpha_{S} $. It is straightforward to verify that
PTN's expressions for the potentials are the same as those of Gupta {\em et
al.} to the second order in $\alpha_{S} $ by using Eq. (\ref{tr}) to connect
the two renormalization schemes. The signs of the terms in Eq. (\ref{tr})
indicate that we should expect smaller coupling constants in the present
calculation than in our 1991 calculation.

In determining the matrix elements for the perturbation improvement of the
central potential energies, we must exercise some care since the total
spin, $ {\bf S} = {\bf S}_{1} + {\bf S}_{2} $, is not a good quantum number
in the most general case. This is a consequence of the fact that its 
magnitude does not commute with the spin-dependent potential because,
for example,

\begin{equation}
\left[ {\bf  L} \cdot {\bf S_{\rm 1} }, S^{2} \right] = 2 i {\bf L} \cdot
     ({\bf S_{2}} \times {\bf S_{1}} ).
\end{equation}	
We will deal with the more complicated case of the mass-mixing matrix in the
next section. For the remainder of this section, we will consider only the
diagonal matrix elements. This will suffice to determine the
requisite expectation values for the case of equal masses, since then the
total spin is a good quantum number.

The diagonal matrix elements can all be evaluated in terms of the expectation
values \cite{ei94} of $\left< {\bf L} \cdot {\bf S} \right> $ since

\begin{equation}
\left< {\bf L} \cdot {\bf S}_{1} \right> \; = \; \left< {\bf L} \cdot
    {\bf S}_{2} \right> \; = \; \frac{1}{2} \left< {\bf L} \cdot {\bf S}
     \right>,
\end{equation}
and
\begin{equation}
\left< S_{12} \right> = \frac{ 4 \left< L^{2} \right> \left< S^{2} \right>
      \; - \; 6 \left< {\bf L } \cdot {\bf S} \right> \; - \;
       12 \left< {\bf L} \cdot {\bf S} \right>^{2} }{ (2{\it l} + 3)
       ( 2{\it l} - 1 ) },
\end{equation}
where {\it l} is the orbital angular momentum quantum number,
as we have verified \cite{la65,kw88,lu89}. Hence, the diagonal matrix
elements of the spin-dependent potentials may be expressed

\begin{equation}
\label{sddg}
\left< V_{SD} \right> = \left< {\bf L} \cdot {\bf S} \right> \left[
              \frac{ \left< T_{a} \right> }{ 4 m_{1}^{2} } +
              \frac{ \left< T_{a}^{\prime } \right> }{4 m_{2}^{2} } +
              \frac{ \left< T_{b} \right> }{ m_{1} m_{2} } \right]
              + \frac{1}{m_{1} m_{2}}
   \left[ \left< S_{12} \right> \: \left< T_{c} \right> + \left< {\bf S}_{1}
   \cdot {\bf S}_{2} \right> \: \left< T_{d} \right> \right]. 
\end{equation}

Our results for the charmonium and upsilon energies obtained with Richardson's
potential and the one-loop expressions for the spin-dependent potentials 
are presented in Tables \ref{tab2} and \ref{tab3} 
(listed in column 4 as FUII98), where they are compared with
the results of Eichten and Quigg \cite{ei94}. The flavor-independent potential
parameters and constituent masses used to obtain these results are

\begin{mathletters}
\label{pa}
\begin{equation}
A=0.152 \: {\rm GeV^{2}}, \hspace{0.2in} \Lambda = 0.431 \: {\rm GeV},
\end{equation}
\begin{equation}
m_{b} =4.889 \: {\rm GeV }, \hspace{0.2in}
m_{c} =1.476 \: {\rm GeV}.
\end{equation}
\end{mathletters}
For the upsilon system the value of the coupling constant $\alpha_{\bar S }
= 0.30, $ and the value of the renormalization scale $\mu = 1.95 \: {\rm GeV}$.
For charmonium these values are $\alpha_{\bar S } = 0.486$ and $\mu = 0.80 \:
{\rm GeV} $. In both cases the value of the universal QCD scale determined
from
\begin{equation}
\Lambda_{QCD} = \mu e^{ - 6 \pi / (33 - 2n_{f}) \alpha_{\bar{S}} },
\end{equation}
is 0.190 GeV, a value consistent with other determinations \cite{pa96}.
Agreement between the calculated results and the measured values \cite{pa96}
in Tables \ref{tab2} and \ref{tab3} is extremely good, the average deviation
being only
4.3 MeV for the upsilon system. Most of the 19.9 MeV deviation in the
charmonium case arises from a 25 MeV discrepancy in the
center-of-gravity of the 1P state. Much of this difference could probably 
be removed by a proper account of the spin-independent relativistic
corrections \cite{br97}.

Our results for the upsilon and charmonium leptonic widths are presented
in Table \ref{tab4}, where they are compared with those calculated by
Eichten and Quigg \cite{ei94}. Our decay rates were obtained with the
formula \cite{cl79},

\begin{equation}
\label{le}
\Gamma_{ee} = \frac{4 \alpha^{2} e_{Q}^{2}}{M^{2}(Q \bar{Q} )}
           |R(0)|^{2} \left[ 1 - \frac{16 \alpha_{S} } {3 \pi } \right],
\end{equation}
where $\alpha $ is the fine structure constant, $e_{Q}$ denotes the quark
charge, and M denotes the mass of the quark-antiquark state. Since the average
values of the momentum transferred in the annihilation graphs underlying
Eq. (\ref{le}) should be much larger than that associated with the scattering
processes of the spin-dependent potentials above, we should expect the
appropriate value for $\alpha_{S}$ in Eq. (\ref{le}) to be smaller than
that used above for the fine structure splittings. Thus we choose
$\alpha_{S} = 0.18$, a value obtained from heavy quarkonium decays
\cite{fu91,kw882}. Our leptonic widths also agree with experiment very well.
In order to carry out a fair comparison with
the widths of Eichten and Quigg, it is necessary
to correct their published results for radiative QCD corrections. Doing
this as described above gives the EQ results shown in parentheses and moves
their results much closer to the measured values. The differences between
the values listed in columns 2 and 3 of Table \ref{tab4} gives a measure of
the differences in wave functions calculated with Richardson's potential and
the Buchm\"{u}ller-Tye potential.

In order to have a good basis for comparison, we have carried out a
Richardson's potential calculation with tree-level expressions for all
the spin-dependent potentials. These may be readily obtained by omitting
all terms of $O(\alpha_{\bar{S}}^{2}) $ in the spin-dependent potentials
above, and thus there is no reference to the renormalization scale.
Then the expressions for the functions T are especially simple, that is,

\begin{mathletters}
\begin{equation}
T_{a} = T_{a}^{\prime } = \frac{ 4 \alpha_{\bar{S}}}{3 r^{3}} - \frac{A}{r},
\end{equation}
\begin{equation}
T_{b} = \frac{4 \alpha_{\bar{S}}}{3 r^{3}},\hspace{0.2in} T_{b}^{\prime } =0,
    \hspace{0.2in} T_{c} = \frac{\alpha_{\bar{S}}}{3 r^{3}}, \hspace{0.2in}
     T_{d} = \frac{32 \pi \alpha_{\bar{S}}} {9} \delta^{3} ({\bf r}).
\end{equation}
\end{mathletters}
The central potential parameters in this calculation are the same as those
listed in Eq. (\ref{pa}). The constituent mass for the charmed quark is
1.476 GeV, the same as in Eq. (\ref{pa}), and $m_{b} = 4.884 \: {\rm GeV}$, 
slightly smaller. The coupling constant $\alpha_{S} = 0.339 $. Results of
this calculation are also presented in Tables \ref{tab2} and \ref{tab3}
(designated as FUI98 in column 3). For both charmonium and the upsilon
system, the average deviation is smaller for the one-loop calculation, but
this does not tell the whole story. The fit to the fine structure splittings
of charmonium with the one-loop calculation is much better than either of the
tree-level calculations presented in Table \ref{tab2}. This is our basis for the expectation that a one-loop calculation should give a more accurate rendering
of the fine structure and the hyperfine structure of the $B_{c}$ system.
Since the central potential parameters are the same, the leptonic widths
for the FUI and FUII calculations are the same.

One of the most dramatic differences in the one-loop and tree level 
results is in the prediction for upsilon hyperfine splittings in
Table \ref{tab3}. For example, the one-loop prediction for the 1S
state is 55 MeV, which is consistent with earlier predictions \cite{fu91,gu82}
and the average of the tree-level predictions is 90 MeV. It is
interesting to note that the one-loop level prediction is much closer to the
lattice results (44-50 MeV) of Bali, Schilling and Wachter \cite{ba97}.
Thus, measurement of the energies of the singlet S energies would serve to
clarify the proper input for a good phenomenological calculation of the
properties of heavy quark systems. Some experimental group should give these 
measurements a high priority.

\section{mass mixing matrix for P states}
\label{sec:ma}
In determining the eigenvalues and eigenfunctions for the $B_{c}$ system,
we have a choice of using basis functions from either the L-S coupling 
scheme or the j-j coupling scheme. We follow the lead
of Eichten and Quigg \cite{ei94} and choose the j-j basis. First we
form the total angular momentum of the charmed quark, ${\bf J}_{1} = {\bf L} +
{\bf S}_{1} $, and then we form the total angular momentum of the system,
${\bf J} = {\bf J}_{1} + {\bf S}_{2}.$ For P states, the $ J = 2 $ states
and the $ J = 0 $ state are the same in either basis, that is,

\begin{equation}
\psi_{2m} (\frac{3}{2} \: \frac{1}{2}) = \psi_{2m} (1 \: 1), \hspace{0.2in}
\psi_{00} (\frac{1}{2} \: \frac{1}{2}) = \psi_{00} (1 \: 1),
\end{equation}
and the determination of the eigenvalues of the spin-dependent potentials
requires only the diagonal elements of Eq. (\ref{sddg}). On the other hand,
the $ J = 1 $ states are a combination,

\begin{equation}
\psi_{1m} = a_{1} \psi_{1m}(\frac{3}{2} \: \frac{1}{2}) + a_{2}
         \psi_{1m} (\frac{1}{2} \: \frac{1}{2}),
\end{equation}
and one must diagonalize the mass mixing matrix in order to determine the
eigenvalues and eigenfunctions. In the basis described above, we have 
derived the following forms for the
spin-dependent operators of Eq. (\ref{sd}), that is,

\begin{mathletters}
\label{mat}
\begin{equation}
\label{ls12}
\left( {\bf L} \cdot {\bf S}_{1} \right) = \left( \begin{array}{cc}
 \frac{1}{2} & 0 \\
\vspace{0.1in}
 0 & -1 \\
\end{array}
\right), \hspace{0.3in} \left( {\bf L} \cdot {\bf S}_{2} \right) =
\left( \begin{array}{cc}
 -\frac{5}{6} & -\frac{\sqrt{2}}{3} \\
\vspace{0.1in}
-\frac{\sqrt{2}}{3} & \frac{1}{3} \\
\end{array} \right),
\end{equation}
and
\begin{equation}
\left( S_{12} \right) = \left( \begin{array}{cc}
\frac{2}{3} & \frac{2 \sqrt{2}}{3} \\
\vspace{0.1in}
\frac{2 \sqrt{2}}{3} & \frac{4}{3}
\end{array} \right), \hspace{0.3in}
\left( {\bf S}_{1} \cdot {\bf S}_{2} \right) = \left( \begin{array}{cc}
- \frac{5}{12} & \frac{\sqrt{2}}{3} \\
\vspace{0.1in}
 \frac{\sqrt{2}}{3} & - \frac{1}{12} \\
\end{array} \right).
\end{equation}
\end{mathletters}
To obtain the matrices for the operators ${\bf L} \cdot {\bf S}$ and
${\bf L} \cdot ({\bf S}_{1} - {\bf S}_{2}) $, one simply takes the 
appropriate linear combination of the matrices in Eqs. (\ref{ls12}). By 
inspection one can
verify that our first 3 matrices in Eqs. (\ref{mat}) are consistent with
Eqs. (2.21) of Eichten and Quigg \cite{ei94}.

It is straightforward to take the equal-mass, or L-S, limit. Then one can
find a unitary transformation that simultaneously diagonalizes the operators,
${\bf L} \cdot {\bf S}$, $S_{12}$ and ${\bf S}_{1} \cdot {\bf S}_{2} $. The
eigenvectors are given by

\begin{mathletters}
\label{lslim}
\begin{equation}
\psi_{1m} (^{3} \! P_{1}) = \sqrt{\frac{1}{3}} \psi_{1m} (\frac{3}{2} \:
   \frac{1}{2})  + \sqrt{\frac{2}{3}} \psi_{1m} (\frac{1}{2} \: \frac{1}{2}),
\end{equation}
\begin{equation}
\psi_{1m} (^{1} \! P_{1}) = \sqrt{\frac{2}{3}} \psi_{1m} (\frac{3}{2} \:
  \frac{1}{2}) - \sqrt{\frac{1}{3}} \psi_{1m} (\frac{1}{2} \: \frac{1}{2}).
\end{equation}
\end{mathletters}
The first eigenvector of Eqs. (\ref{lslim}) has the lower eigenvalue
since it is determined by choosing the solution to the quadratic 
eigenvalue equation with the minus sign. Our choice of overall sign for
the second eigenvector of Eqs. (\ref{lslim}) is opposite that of
Eichten and Quigg, but the same as that of Gershtein {\em et al. }
\cite{ge95}.

\section{results and discussion}
\label{sec:re}
Since the central potential parameters are strictly flavor independent and
the constituent masses are not allowed to vary, the only decision that one has
to make to address the $B_{c}$ system is to decide on the value of
$\alpha_{\bar{S} }$. The simplest choice is the average of the values in the 
charmonium and the upsilon system, that is, $\alpha_{\bar{S} } = 0.393 $.
This choice requires a value of $\mu = 1.12 \; {\rm GeV} $, to preserve 
the value of 0.190 GeV for the QCD scale.
Our results for the low-lying S, P and D states of the $B_{c}$ system
are presented in column 6 of 
Table \ref{tab5}, where they are compared with the predictions of Eichten and
Quigg \cite{ei94}, Gershtein {\em et al.} \cite{ge95}, Chen and Kuang
\cite{ch92}, the NRQCD lattice calculation \cite{da96} and
our tree level calculation. The energy of each of these states lies below
the B-D meson threshold at 7143 MeV.  In determining the D-state
splittings, which are much smaller than the P-state splittings, we followed the
example of Eichten and Quigg and did not use the mass-mixing matrix approach.
Several running coupling constant effects are noticeable in Table \ref{tab5},
although our results are in general agreement with the earlier calculations.
Our S-state hyperfine splittings are smaller than those of Eichten and Quigg,
and our results for the P-state fine structure splittings are larger. We do
not see the inverted order of the D states predicted by both Eichten and Quigg
and Gershtein {\em et al.}. The order of the fine-structure levels 
\cite{sc82,is98} is
a consequence of the relative sizes of the the perturbative and the
nonperturbative contributions to the spin-orbit potentials of Eq. (\ref{sd}).

Comparison with the lattice NRQCD results \cite{da96} is of special    
interest. Their result for the $B_{c}$ mass is based on a lattice
calculation of the kinetic mass, $4.76 \pm 0.02$ and a determination of
the scale factor by comparison with charmonium and the upsilon system,
which gives  $ a^{-1} = 1.32\pm 0.04\; {\rm GeV}$. Thus, we list two errors for
the lattice calculation of the $B_{c}$ mass in Table \ref{tab5}. The smaller
error is a consequence of the error in the kinetic mass calculation and the
larger error is a consequence of the uncertainty in the overall scale. In
determining the other lattice errors, we simply list the largest possible
source of error without including the error due to the overall scale. The
lattice result for the 1S hyperfine splitting is $41 \pm 3 \; {\rm MeV}$, which
is independent of the larger errors discussed above, and somewhat smaller
than our 55 MeV result. The lattice result for the $1P^{2} - 1P^{0}$ splitting
is $59 \pm 5 \; {\rm MeV}$, in reasonable agreement with our result of
71 MeV. Although the number quoted in Ref. \cite{da96} for the 
$B_{c}$ mass is very close to our result, its importance as a confirmation
of our work is undermined by the large error in the overall scale. Another
important lattice result is now available from the UKQCD 
collaboration \cite{uk99}.
Their result for the ground state energy is 
$6386 \pm 9 \pm 98 \pm 15 $ MeV,
which is consistent with our result because of the large error bars. 
It will be interesting to see whether further refinements of the lattice
calculations will support our results, or offer the experimenters alternative
predictions.

In order to get some idea of an error estimate for our predictions, we have
calculated the ground state mass of the $B_{c}$ system as a function of
$\alpha_{\bar{S}} $ in a range bounded by the values determined in
charmonium and the upsilon system. Our results are shown in Fig. \ref{fig1},
where they are compared with the result of Table \ref{tab1}, the 
Eichten-Quigg quote and the NRQCD lattice result. The error shown for the 
lattice result simply ignores the large error associated with the overall
scale. The horizontal lines there
describe the limits determined by Kwong and Rosner \cite{kw91}.
Using the largest and smallest values of $\alpha_{\bar{S}}$ in Fig. \ref{fig1}
to determine the errors, we have

\begin{equation}
\label{prm}
M_{B_{c}} = 6286_{-6}^{+15} \; {\rm MeV} \hspace{0.4in} M_{B_{c}^{*}} =
      6341_{-5}^{+2} \; {\rm MeV},
\end{equation}
as our predicted value for the 1S energies. It is interesting to note
how close our results are to the earlier predictions of Godfrey and
Isgur \cite{go85} (6270 and 6340 MeV) and the predictions of
Baker, Ball and Zachariasen \cite{ba92} (6287 and 6372 MeV), although
our prediction for the hyperfine splitting is smaller than either.
Clearly, the precision of the experiments \cite{ab98}
requires a very substantial
improvement to be sensitive to the energy differences between the
various calculations listed in Table \ref{tab5} and Fig. \ref{fig1}.

The pseudoscalar decay constant is given by the Van Royen-Weisskopf
formula modified for color \cite{ro67}, that is,

\begin{equation}
f_{B_{c}}^{2} = \frac{ 3 |R_{1S}(0)|^{2} }{\pi M_{B_{c}} },
\end{equation}
and we find that

\begin{equation}
\label{prf}
f_{B_{c}} = 517 \; {\rm MeV},
\end{equation}
in excellent agreement with Eichten and Quigg's result, $f_{B_{c}} = 500
\; {\rm MeV} $, and in reasonable agreement with the lattice result \cite{da96},
$ f_{B_{c}} = 440 \pm 20 \; {\rm MeV} $. 

The empirical result obtained
by Collins {\em et al.} \cite{co97} for potential model wave functions at
the origin, that is,
\begin{equation}
|R_{B_{c}}(0)|^{2} \cong |R_{J/\psi}(0)|^{1.3} |R_{\Upsilon}(0)|^{0.7},
\end{equation}
provides another touchstone for our numerical work. Using this relationship
and input from our charmonium and upsilon calculations,
we get $|R_{B_{c}}(0)|^{2} \cong 1.81 \; {\rm GeV}^{3} $, about 3\% higher 
than our numerical
result. Using their empirical relationship for the ground state hyperfine
splittings, that is,
\begin{equation}
M_{B_{c}^{*}} - M_{B_{c}} \simeq 0.7 \left(M_{J/\psi} - M_{\eta_{c}} 
	\right)^{0.65} \left( M_{\Upsilon } - M_{\eta_{b}} \right)^{0.35},
\end{equation}
yields a splitting of 63 MeV, about 14\% larger than our result listed in
Table V. Both of these results are reasonable in view of the spread of the
results Collins {\em et al.} obtained with different forms for the central
potential.

Diagonalizing the P-state mixing matrix, we obtain the following
combinations for the two lowest $J = 1$ P states,

\begin{mathletters}
\label{mix}
\begin{equation}
\psi_{1m} (1^{+}) = 0.118 \:  \psi_{1m} ( \frac{3}{2} \: \frac{1}{2} )
                  + 0.993 \: \psi_{1m} ( \frac{1}{2} \: \frac{1}{2} ),
\end{equation}
\begin{equation}
\psi_{1m} (1^{+'}) = 0.993 \:  \psi_{1m} ( \frac{3}{2} \: \frac{1}{2} )
                   - 0.118 \: \psi_{1m} ( \frac{1}{2} \: \frac{1}{2} ),
\end{equation}
\end{mathletters}
which is very close to the j-j coupling limit, that one would expect to be
valid in the heavy-quark limit.  Using the inverse of
Eqs. (\ref{lslim}), we can determine the probability of observing
spin 1 in the lowest $1^{+}$ state, $P_{1^{+}} (S=1) = 0.773$. Our result
is consistent with the lattice calculation \cite{da96}, where
the mixing angle in the L-S basis was found to be close to that of the
j-j limit. From this mixing angle ($\theta = 33.4^{o} \pm 1.5^{o}$)
we obtain $P_{1^{+}} (S=1) = 0.697 \pm 0.020 $, in reasonable agreement 
with our result. Our results for the mixing angles of the $1^{+}$ and $1^{+'}$
states is very different from that of Eichten and Quigg, whose results
were much closer to the L-S limit. Below we show that this difference has
important implications for the spectrum of photons emitted in electric
dipole transitions.

The electric dipole rate for the emission of a photon \cite{ei94} of energy
k is given by

\begin{equation}
\label{dip}
\Gamma_{E1}(i \rightarrow f \: + \: \gamma ) = \frac{ 4 \alpha \left<
      e_{Q} \right>^{2} }{27} k^{3} (2 J_{f} + 1 ) |\left< f | r | i
      \right> |^{2} \: S_{if},
\end{equation}
where the statistical factor $S_{if} = 1 $ for transitions between
triplet S and triplet P states \cite{ei77}, and $S_{if} = 3 $ for transitions
between spin-singlet states. The mean charge in Eq. (\ref{dip}) is

\begin{equation} \left< e_{Q} \right> = \frac{ m_{2} e_{1} - m_{1} e_{2} }{m_{1} + m_{2}}, 
\end{equation}
where $e_{1}$ denotes the charge of the charmed quark (in units of the proton's
charge) and $e_{2}$ denotes the charge of the bottom antiquark. Our results for the $ 1P \rightarrow 1S $ and the $ 2S \rightarrow 1P $ transition
rates are shown in Table \ref{tab6}, where they are compared with Eichten and 
Quigg's results. Our rates for transitions involving the
$1P^{2}$ and $1P^{0}$ states are rather close to their counterparts 
calculated by Eichten and Quigg. However for the $J=1$ states important
differences arise. In particular, we predict 4 nonzero transition
probabilities instead of 2. Thus, each of our simulated photon spectra presented
in Fig. \ref{fig2} and Fig. \ref{fig3} has six lines instead of 4. These
additional lines are a consequence of the fact that our mixture of states is
in Eq. (\ref{mix}) is not close to the L-S limit, and thus both $1^{+}$ states
have a substantial admixture of both triplet and singlet components.

The magnetic dipole transition rate between S states is given by

\begin{equation}
\Gamma_{M1}(i \: \rightarrow \: f + \gamma) = \frac{16 \alpha}{3} 
\mu_{maq}^{2} k^{3} (2 J_{f} + 1) |\left< f | j_{0}(kr/2) | i \right> |^{2},
\end{equation}
where the mean magnetic dipole moment is

\begin{equation}
\mu_{mag} = \frac{m_{2} e_{1} - m_{1} e_{2} } { 4 m_{1} m_{2} }.
\end{equation}
Our results for the magnetic dipole transition rates are presented in Table
\ref{tab7}, where they are compared with those of Eichten and Quigg \cite{ei94}
and Gershtein {\em et al.} \cite{ge95}. Most of the differences are a 
consequence of different  energies for the hyperfine splittings,
since the results for the
matrix elements are rather close.  

The photon energies and transition rates in Tables \ref{tab6} and \ref{tab7}
suggest at least two good strategies for experimental searches for the
1P and 1S states. One could detect one or both of the high energy photons
(457, 436 MeV) associated with the decay of the lowest two $1^{+}$ states to the ground state and then search for a leptonic decay of the ground state. An
alternative would be to look for some of the high-energy photons (417, 406,
 384, 350 MeV) in the $ 1P \rightarrow 1^{3} \! S_{1} $ transitions and
then seek a coincidence with the 55 MeV photon associated with the decay to
the ground state.

Since the charmed quark and the bottom antiquark cannot annihilate into
gluons, the only additional complication that arises in the decays of
these low-lying states is the $\pi \: \pi $ channel.
If we take the rates for the $\pi \: \pi $ decays from Eichten
and Quigg, then we can work out complete decay schemes and branching
ratios for the 2S and 1P states as well as the $1^{3} \! S_{1} $ state.
These are shown in Table \ref{tab8}. Our table of decay rates and branching
ratios differs from that of Eichten and Quigg in two important respects.
More photon channels are available to the $1^{+}$ states and the 2S states.
Our decay width for the $1^{3}
\!S_{1}$ is 59 eV, more than a factor of 2 smaller than theirs.

Our final calculation is that of the lifetime of the ground state of the
$B_{c} $ system. We follow the approach used by several 
researchers \cite{ge91,bi96,el98} where the decay of the $B_{c} $ meson is 
taken to be 
the sum of 3 distinct contributions, namely weak decay of the $\bar{b} $
antiquark while the {\em c} quark behaves as a spectator, weak decay of the
{\em c} quark while the $\bar{b} $ antiquark behaves as a spectator and an
annihilation of the $\bar{b} $ antiquark and the $c$ quark into an intermediate
vector boson that subsequently decays into a lepton-neutrino pair or a 
quark-antiquark pair. Thus the total decay rate is the sum,
\begin{equation}
\Gamma (B_{c} \rightarrow X) = \Gamma (\bar{b} \rightarrow X) +
	\Gamma (c \rightarrow X) + \Gamma (annih).
\end{equation}
If one neglects quark binding effects, then the first two terms are given by
\begin{equation}
\label{gambc}
\Gamma (\bar{b} \rightarrow X) = \frac{9 G_{F}^{2} |V_{cb}|^{2} m_{b}^{5} }
	{192 \pi^{3}} \; , \; \Gamma (c \rightarrow X) = \frac{ 5 G_{F}^{2}
	|V_{cs}|^{2} m_{c}^{5}}{192 \pi^{3}},
\end{equation}
where the subscripted quantities {\em V} denote the appropriate 
Cabibbo-Kobayashi-Maskawa matrix elements \cite{ppb98} and $G_{F} $ denotes
the Fermi coupling constant. Using the constituent masses listed in 
Eq. (\ref{pa}), we obtain
\begin{equation}
\Gamma (\bar{b} \rightarrow X) = \left( 8.73 \pm 1.34 \right) \times 10^{-10}\;
	{\rm MeV} \; , \; \Gamma (c \rightarrow X) = \left( 7.59 \pm
	0.02 \right) \times 10^{-10} \; {\rm MeV}.
\end{equation}
The annihilation width is given by
\begin{equation}
\Gamma (annih) = \frac{ G_{F}^{2} }{8 \pi } |V_{bc}|^{2} f_{B_{c}}^{2} M_{B_{c}}	\sum_{i} m_{i}^{2} \left( 1 - \frac{m_{i}^{2}}{M_{B_{c}}^{2}} 
	\right)^{2} C_{i},
\end{equation}
where $m_{i} $ denotes the mass of the heavier Fermion in the given decay
channel.
The most important channels in the sum are the $\tau \nu $ and the $\bar{c} s$ 
channels. For the former $ C_{i} = 1 $, and for the latter 
$C_{i} = 3 |V_{cs}|^{2}. $ 
Using Eq. (\ref{prf}) for $f_{B_{c}} $ and Eq. (\ref{prm}) for
$M_{B_{c}} $, we have
\begin{equation}
\Gamma (annih) = \left( 1.13 \pm 0.17 \right) \times 10^{-10} \; {\rm MeV}.
\end{equation}
Adding these 3 widths yields a lifetime
\begin{equation}
\tau_{B_{c}} = 0.38 \pm 0.03 \; {ps},
\end{equation}
in good agreement with the measured CDF result \cite{ab98}. Our result is also
in reasonable agreement with
the recent calculation of El-Hady, Lodhi and Vary \cite{el98},
who obtain $\tau_{B_{c}} = 0.46 \; {\rm ps} $, since
the spectator widths of Eqs. (\ref{gambc}) are very sensitive to small 
differences in the constituent masses. Although different authors may wish to
interprete relatively small differences between $\Gamma(\bar{b} \rightarrow X)$
and $\Gamma(c \rightarrow X)$ as the domination of one process over the other,
we feel that the safest characterization of our results is that the two
spectator processes are almost equally important and that the annihilation
processes are less important, consistent with the earlier conclusion of 
Gershtein {\em et al. } \cite{ge91}.

\section*{acknowledgements}
The author gratefully acknowledges the hospitality of Professor Werner
Scheid and his colleagues at the Institut f\"{u}r Theoretische Physik
of the Justus Liebig Universit\"{a}t. Support for this work was provided
by the Alexander von Humboldt-Stiftung.

\appendix
\section{Hyperfine splitting and contact potentials}
Determining the energy shifts from the hyperfine splitting                 
can be problematic because of the delta function that is often present
in the spin-spin potential. As Lucha, Sch\"{o}berl and Gromes point out
in their review \cite{lu91}, the energy of the singlet state is not
actually bounded from below, in marked contrast to the first-order perturbation
theory result of Eq. (\ref{hfs}). Although the use of Eq. (\ref{hfs}) in the
literature is fairly common, seldom does one see any discussion of its 
validity. It is straightforward to create a context for addressing this
question, by considering a more general form for the 
spin-spin potential, which allows for a finite range, that is,
\begin{equation}
\label{gauss}
V_{SS} = \frac{32 \alpha_{S}}{9 m_{1} m_{2} } \frac{e^{-r^{2}/b^{2}}}
	{\pi^{1/2} b^{3}} {\bf S}_{1} \cdot {\bf S}_{2}.
\end{equation}
The advantage of such a form is that one can calculate the singlet and
triplet S-state eigenvalues of the Hamiltonian of Eq. (\ref{nrham})
exactly and examine the limit as the range parameter {\em b} becomes
smaller and smaller. We have done such a calculation with the logarithmic
potential of Eq. (\ref{vlog}) and the parameters listed in Table I. Our
results for the exact singlet and triplet as a function of the range {\em b}
are shown in Fig. \ref{fig4}, where they are compared with the results of
a first-order perturbation calculation of the singlet and triplet energies
produced by the potential of Eq. (\ref{gauss}). The contact potential results
of Eq. (\ref{hfs}) are presented as two horizontal lines. It is gratifying
to see the first-order perturbation result from Eq. (\ref{gauss}) above
approach the contact potential result in the limit $b \rightarrow 0 $. 
Extrapolating the exact result for the triplet energy to the $b = 0 $
limit gives 6330 MeV, about 3 MeV lower than the result listed in column 4
of Table 1, which was obtained with the contact potential. Thus, it is clear
that first-order perturbation theory and the contact potential
give a good account of the 1S triplet energy.

Fig. \ref{fig4} gives a clear signal of the instability of the singlet 
energies as $b \rightarrow 0 $. However the difficulty begins to
appear only  as {\em b} decreases below 0.3 ${\rm GeV}^{-1} $. The short range 
required for the effects of the instability to manifest itself provides
a means of resolving this dilemma. As Lucha, Sch\"{o}berl and Gromes point out,
the contact potential expression is not really valid for such short ranges
since one must take the nonrelativistic limit in order to obtain it. Such
a limit requires that nonlocal effects associated with the normalization of
Dirac wave functions be ignored. Thus it might be reasonable to choose 
$b= 0.36 \; GeV $, the geometric mean of the Compton wavelengths of the
two constituent quarks. Such a choice would lead to a value of the 
singlet energy very close to that found in column 4 of Table 1.

\begin{figure}
\caption{Ground state energy of the $B_{c}$ system as a function of the
running coupling constant. The PH98 result is based on the calculations
presented in Table \ref{tab1}.}
\label{fig1}
\end{figure}

\begin{figure}
\caption{Simulated photon spectrum for $1P \rightarrow 1S$ transitions.
         The probability of populating each of the initial states is assumed
         to be equal.}
\label{fig2}
\end{figure}

\begin{figure}
\caption{Simulated photon spectrum for $2S \rightarrow 1P$ transitions.}
\label{fig3}
\end{figure}

\begin{figure}
\caption{Hyperfine splittings of the 1S state for a finite range spin-spin
	potential.}
\label{fig4}
\end{figure}

\newpage

\begin{table}
\caption{Ground state energies of the heavy-quark systems (MeV).}
\label{tab1}
\begin{tabular}{lccccc}

State & Cornell & Mart\'{\i}n & logarithm & Song-Lin & Turin \\
\tableline
$\alpha_{S} $ & 0.313 & 0.437 & 0.372 & 0.396 & 0.373 \\
$m_{b} (GeV)$ & 5.232 & 5.174 & 4.905 & 5.199 & 5.171 \\
$m_{c} (GeV)$ & 1.840 & 1.800 & 1.500 & 1.820 & 1.790 \\
$1^{3}\!S_{1} (c \bar{c})$ & 3097 & 3097 & 3097 & 3097 & 3097 \\
$1^{1}\!S_{0} $ & 2980 & 2980 & 2980 & 2980 & 2980 \\
$\Delta E_{1S}$ & 117 & 117 & 117 & 117 & 117 \\
$1^{3}\!S_{1} (b \bar{b})$ & 9461 & 9461 & 9460 & 9460 & 9460 \\
$1^{1}\!S_{0} $ & 9316 & 9397 & 9395 & 9380 & 9365 \\
$\Delta E_{1S}$ & 145 & 64 & 65 & 80 & 95 \\
$1^{3}\!S_{1} ( c\bar{b} )$ & 6337 & 6319 & 6333 & 6324 & 6327 \\
$1^{1}\!S_{0} $ & 6246 & 6247 & 6266 & 6247 & 6247 \\
$\Delta E_{1S} $ & 91 & 72 & 67 & 77 & 80 \\
\end{tabular}
\end{table}

\begin{table}
\caption{Charmonium energies (MeV).}
\label{tab2}
\begin{tabular}{ccccc}

State & EQ94 & FUI98 & FUII98 & EXPT \\
\tableline
$1^{3}\!S_{1} $ & 3097 & 3098 & 3098 & $3097 \: \pm \: 0.04 $ \\
$1^{1}\!S_{0} $ & 2980 & 2981 & 2980 & $ 2980 \: \pm \: 2 $ \\
$2^{3}\!S_{1} $ & 3686 & 3692 & 3693 & $ 3686 \: \pm \: 0.1 $ \\
$2^{1}\!S_{0} $ & 3608 & 3617 & 3615 &   \\
$1^{3}\!P_{2} $ & 3507 & 3515 & 3530 & $ 3556 \: \pm \: 0.1 $ \\
$1^{3}\!P_{1} $ & 3486 & 3492 & 3482 & $ 3511 \: \pm \: 0.1 $ \\
$1^{3}\!P_{0} $ & 3436 & 3443 & 3391 & $ 3415 \: \pm \: 1.0 $ \\
$1^{1}\!P_{1} $ & 3493 & 3499 & 3501 & $ 3526 \: \pm \: 0.1 $ \\
$\sqrt{\delta^{2}} $ & 25.5 & 22.7 & 19.9 &  \\
\end{tabular}
\end{table}

\begin{table}
\caption{Upsilon energies (MeV). }
\label{tab3}
\begin{tabular}{ccccc}

State & EQ94 & FUI98 & FUII98 & EXPT \\
\tableline
$1^{3}\!S_{1} $ & 9464 & 9461 & 9461 & $ 9460 \: \pm \: 0.2 $ \\
$1^{1}\!S_{0} $ & 9377 & 9368 & 9406 &  \\
$2^{3}\!S_{1} $ & 10007 & 10022 & 10027 & $ 10023 \: \pm \: 0.3 $ \\
$2^{1}\!S_{0} $ & 9963 & 9978 & 10001 &  \\
$3^{1}\!S_{1} $ & 10339 & 10358 & 10364 & $ 10353 \: \pm \: 0.5 $ \\
$3^{1}\!S_{0} $ & 10298 & 10325 & 10344 & \\
$1^{3}\!P_{2} $ & 9886 & 9902 & 9910 & $ 9913 \: \pm \: 0.6 $ \\
$1^{3}\!P_{1} $ & 9864 & 9881 & 9891 & $9892 \: \pm \: 0.7 $ \\
$1^{3}\!P_{0} $ & 9834 & 9852 & 9863 & $ 9860 \: \pm \: 1 $ \\
$1^{1}\!P_{1} $ & 9873 & 9889 & 9899 & \\
$2^{3}\!P_{2} $ & 10242 & 10261 & 10269 & $ 10269 \: \pm \: 0.4 $ \\
$2^{3}\!P_{1} $ & 10224 & 10244 & 10255 & $ 10255 \: \pm \: 0.5 $ \\
$2^{3}\!P_{0} $ & 10199 & 10222 & 10234 & $ 10232 \: \pm \: 0.6 $ \\
$2^{1}\!P_{1} $ & 10231 & 10251 & 10261 & \\
$\sqrt{\delta^{2}} $ & 24.9 & 8.3 & 4.3 & \\
\end{tabular}
\end{table}

\begin{table}
\caption{Leptonic widths (KeV).} 
\label{tab4}
\begin{tabular}{cccc}

State & EQ94 & FU98 & EXPT \\
\tableline
$\Upsilon(1S) $ & $1.71 \: (1.19)$ & 1.34 & $ 1.32 \: \pm \: 0.05 $ \\
$\Upsilon(2S) $ & $0.76 \: (0.53)$ & 0.57 & $ 0.52 \: \pm \: 0.03 $ \\
$\Upsilon(3S) $ & $0.55 \: (0.38)$ & 0.40 & $ 0.48 \: \pm \: 0.06 $ \\
$\psi(1S) $ & $ 8.00 \: (5.55)$ & 5.81 & $ 5.26 \: \pm \: 0.37 $ \\
$\psi(2S) $ & $ 3.67 \: (2.55)$ & 2.61 & $ 2.14 \: \pm \: 0.21 $ \\
\end{tabular}
\end{table}

\begin{table}
\caption{Energies of the $B_{c}$ system (MeV).}
\label{tab5}
\begin{tabular}{lcccccc}

State & EQ94 & GKLT95 & CK92 & FUI98 & FUII98 & LAT96 \\
\tableline
$1^{3}\!S_{1}$ & 6337 & 6317 & 6355 & 6341 & 6341 & $6321 \: \pm \: 30  $\\
$1^{1}\!S_{0}$ & 6264 & 6253 & 6310 & 6267 & 6286 & $6280 \: \pm \: 30 \:  \pm
     190  $\\
$2^{3}\!S_{1}$ & 6899 & 6902 & 6917 & 6911 & 6914 & $6990 \: \pm \: 80 $ \\
$2^{1}\!S_{0}$ & 6856 & 6867 & 6890 & 6869 & 6882 & $6960 \: \pm \: 80 $ \\
$1P_{2}$ & 6747 & 6743 & 6773 & 6761 & 6772 & $6783 \: \pm \: 30 $ \\
$1P \: 1^{+'} $ & 6736 & 6729 &  & 6750 & 6760 & $6765 \: \pm \: 30 $ \\
$1P \: 1^{+} $ & 6730 & 6717 &   & 6742 & 6737 & $6743 \: \pm \: 30  $ \\
$1P_{0}$ & 6700 & 6683 & 6728 & 6713 & 6701 & $6727 \: \pm \: 30  $ \\
$1^{3}\!D_{3}$ & 7005 & 7007 & & 7022 & 7032 & \\
$1^{3}\!D_{2}$ & 7012 &  & & 7025 & 7028 & \\
$1^{3}\!D_{1}$ & 7012 & & & 7024 & 7019 & \\
$1^{1}\!D_{2}$ & 7009 & 7008 &  & 7023 & 7028 &  \\
\end{tabular}
\end{table}

\begin{table}
\caption{Electric dipole matrix elements and transition rates.}
\label{tab6}
\begin{tabular}{lcccccc}

Transition & Photon energy & & $ \left< f | r | i \right> ({\rm GeV })^{-1}$ & 
  & $ \Gamma (KeV) $ & \\
 & (MeV) & GKLT95 & EQ94 & FU98 & EQ94 & FU98 \\
\tableline
$1P_{2} \rightarrow 1^{3}\!S_{1}$ & 417 & & & & 113 & 126 \\
$1P \: 1^{+'} \rightarrow 1^{3}\!S_{1}$ & 406 & & & & 0.1 & 26.2 \\
$1P \: 1^{+} \rightarrow 1^{3} \!S_{1}$ & 384 & & & & 99.5 & 75.8 \\
$1P_{0} \rightarrow 1^{3} \!S_{1} $ & 350 & 1.568 & 1.714 & 1.683 &
           79.2 & 74.2 \\
$1P \: 1^{+'} \rightarrow 1^{1} \!S_{0} $ & 457 & & & & 56.4 & 128 \\
$1P \: 1^{+} \rightarrow 1^{1} \!S_{0} $ & 436 & & & & 0.0 & 32.5 \\
\\
$2^{3} \!S_{1} \rightarrow 1P_{2}$ & 141 & & & & 17.7 & 14.5 \\
$2^{3} \!S_{1} \rightarrow 1P \: 1^{+'}$ & 152 & & & & 0.0 & 2.5 \\
$2^{3} \!S_{1} \rightarrow 1P \: 1^{+}$ & 175 & & & & 14.5 & 13.3 \\
$2^{3} \!S_{1} \rightarrow 1P_{0}$ & 210 & -2.019 & -2.247 & -2.253 &
           7.8 & 9.6 \\
$2^{1} \!S_{0} \rightarrow 1P \: 1^{+'} $ & 121 & & & & 5.2 & 13.1 \\
$2^{1} \!S_{0} \rightarrow 1P \: 1^{+} $ & 143 & & & & 0.0 & 6.4 \\
\end{tabular}
\end{table}

\begin{table}
\caption{Magnetic dipole matrix elements and transition rates.}
\label{tab7}
\begin{tabular}{ccccccc}

Transition & Photon energy & $\left< f | j_{0} ( kr/2) | i \right> $ & &
         & $\Gamma (KeV)$ & \\
 & (MeV) & EQ94 & FU98 & GKLT95 & EQ94 & FU98 \\
\tableline
$2^{3} \!S_{1} \rightarrow 2^{1} \!S_{0}$ & 32 & 0.9990 & 0.9995 & 0.010 &
        0.029 & 0.012 \\
$2^{3} \!S_{1} \rightarrow 1^{1} \!S_{0}$ & 599 & 0.0395 & 0.0399 & 0.098 &
        0.123 & 0.122 \\
$2^{1} \!S_{0} \rightarrow 1^{3} \!S_{1}$ & 520 & 0.0265 & 0.0305 & 0.096 &
        0.093 & 0.139 \\
$1^{3} \!S_{1} \rightarrow 1^{1} \!S_{0}$ & 55 & 9.9993 & 0.9996 & 0.060 &
        0.135 & 0.059 \\
\end{tabular}
\end{table}

\begin{table}
\caption{Decays and branching ratios of the $B_{c}$ system.}
\label{tab8}
\begin{tabular}{lclc}

State & Total width (KeV) & Decay mode & Branching ratio (percent) \\
\tableline
$1^{3} \!S_{1}$ & 0.059 &  $1^{1} \!S_{0} + \gamma $ & 100 \\
 & & & \\
$1P_{2} $ & 126 & $ 1^{3} \!S_{1} + \gamma $ & 100 \\
 & & & \\
$1P \: 1^{+'} $ & 154 & $ 1^{3} \!S_{1} + \gamma $ & 17 \\
  & & $1^{1} \!S_{0} + \gamma $ & 83 \\
 & & & \\
$1P \: 1^{+} $ & 108 & $1^{3} \!S_{1} + \gamma $ & 70 \\
 & & $ 1^{1} \!S_{0} + \gamma $ & 30 \\
 & & & \\
$1P_{0} $ & 74.2 & $1^{3} \!S_{1} + \gamma $ & 100 \\
 & & & \\
$2^{3} \!S_{1} $ & $ 89.9 \pm 7 $ & $1^{3} \!S_{1} + \pi \pi $& $ 56 \pm 8 $ \\
 & & $1P_{2} + \gamma $ & $ 16 \pm 1$ \\
 & & $1P \: 1^{+'} + \gamma $ & $3 \pm 0.2 $ \\
 & & $1P \: 1^{+} + \gamma $ & $15 \pm 1 $ \\
 & & $1P_{0} + \gamma $ & $ 11 \pm 1 $ \\
 & & & \\
$2^{1} \!S_{0} $ & $ 69.5 \pm 7 $ & $1^{1} \!S_{0} + \pi \pi $ & $72\pm 10$ \\
 & & $1P \: 1^{+'} + \gamma $ & $19 \pm 2 $ \\
 & & $1P \: 1^{+} + \gamma $ & $ 9 \pm 1 $ \\
\end{tabular}
\end{table}


\begin{references}
\bibitem{ft1} In the present work we choose $n_{f} = 3 $ to prevent introducing
            further parameters.
\end{references}
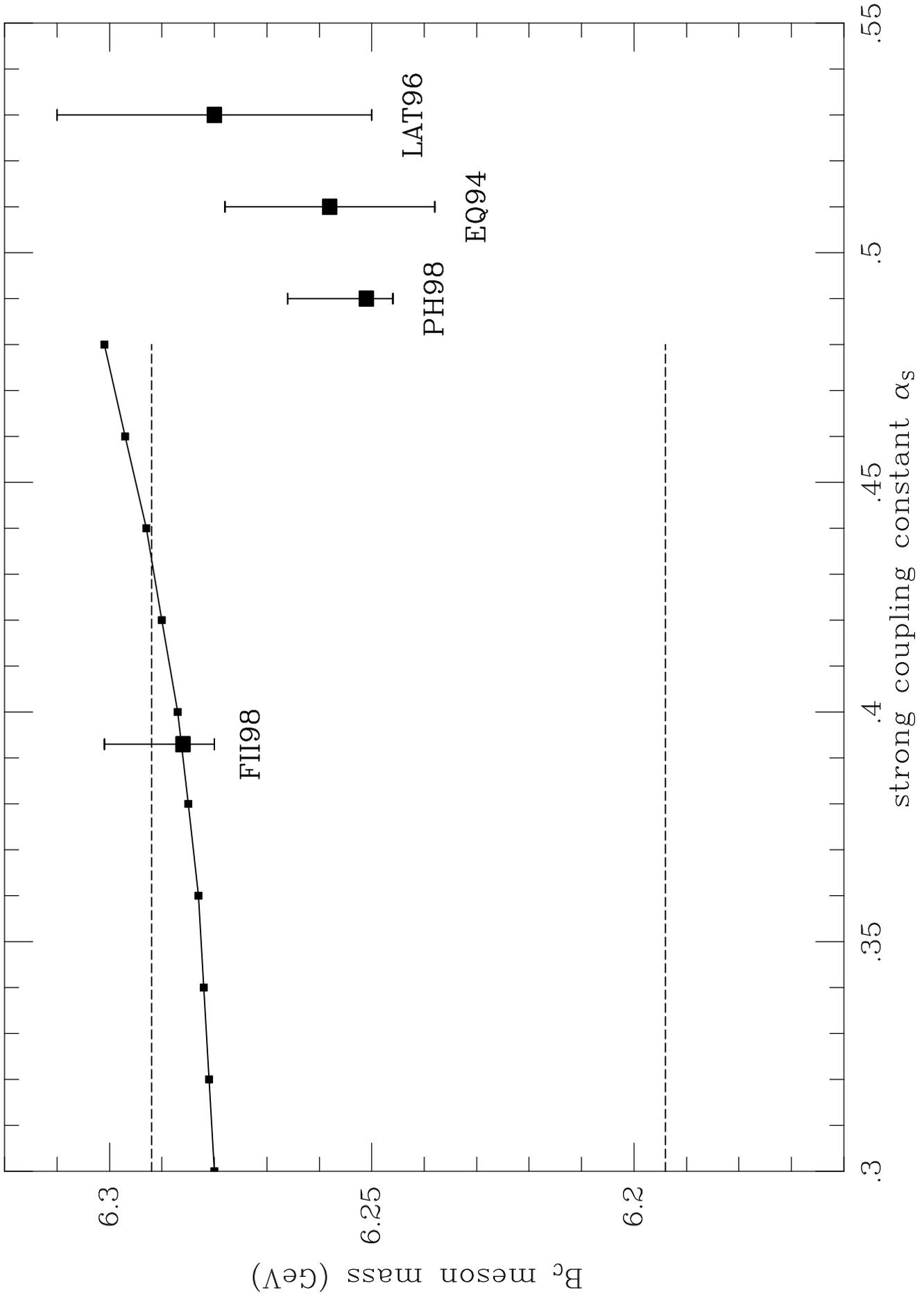
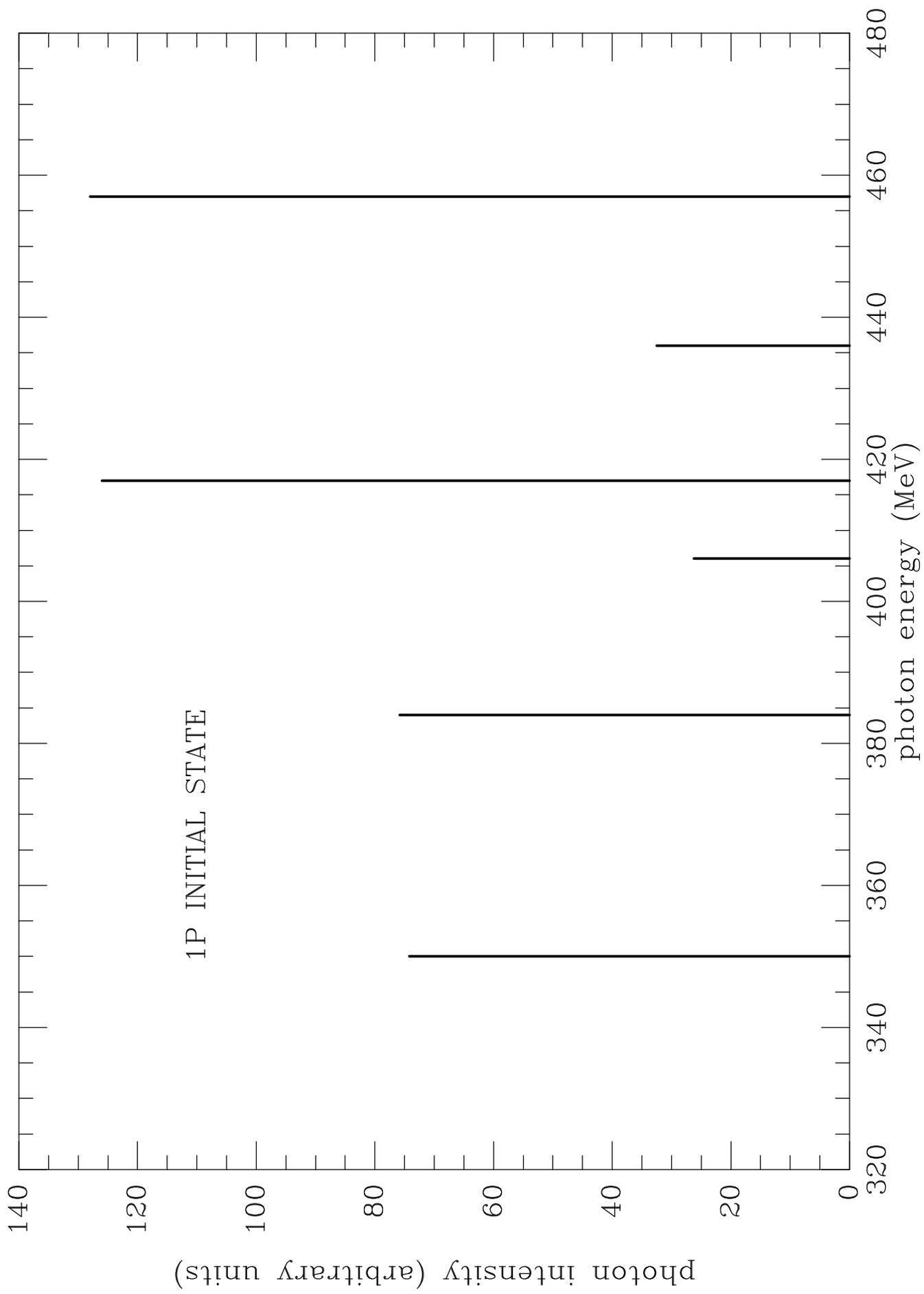
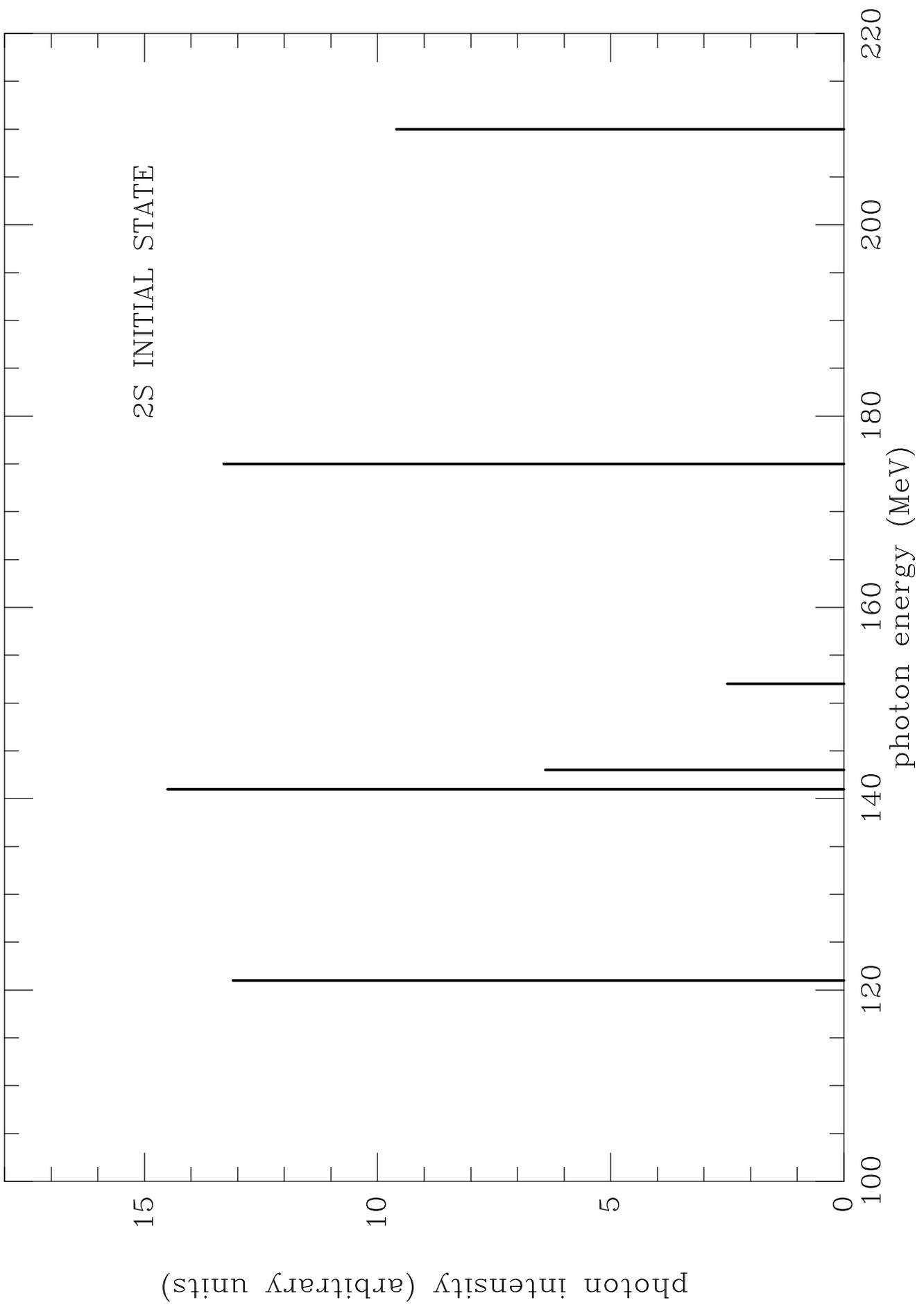
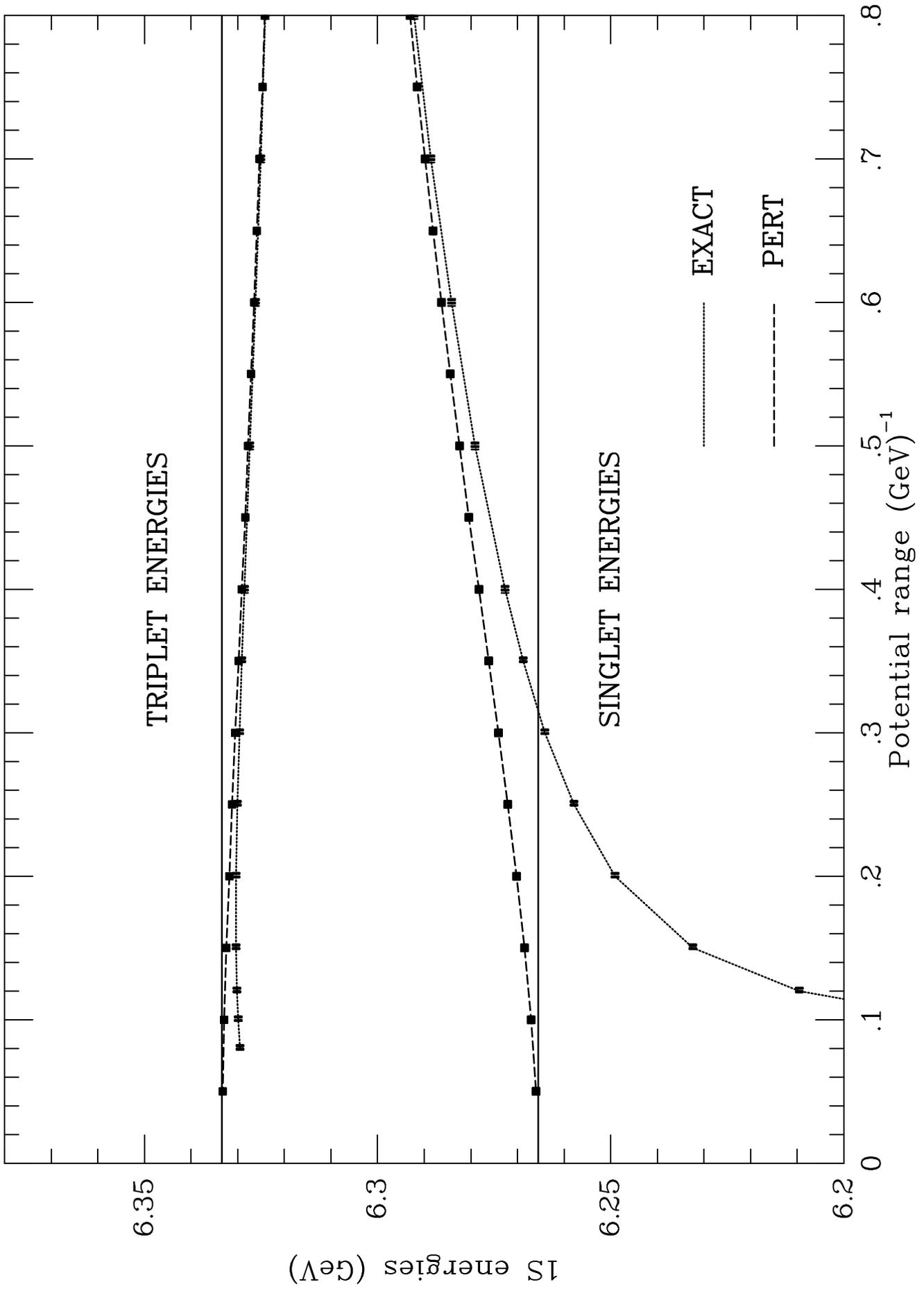
\end{document}